\documentclass[aps,prb,twocolumn,nopacs,superscriptaddress]{revtex4}
\usepackage{graphicx}
\usepackage{verbatim}

\newcommand{\bk}{{\bf k}}

\begin{document}

\title{A Majorana Platform: Spin-orbital groundstates of Superconducting doped topological insulators}

\author{L. Andrew Wray}
\affiliation{Department of Physics, Joseph Henry Laboratories, Princeton University, Princeton, NJ 08544, USA}
\affiliation{Advanced Light Source, Lawrence Berkeley National Laboratory, Berkeley, California 94305, USA}
\author{Yuqi Xia}
\author{Su-Yang Xu}
\affiliation{Department of Physics, Joseph Henry Laboratories, Princeton University, Princeton, NJ 08544, USA}
\author{Dong Qian}
\affiliation{Department of Physics, Joseph Henry Laboratories, Princeton University, Princeton, NJ 08544, USA}
\affiliation{Department of Physics, Shanghai Jiao-Tong University, Shanghai 200030, People's Republic of China}
\author{Alexei V. Fedorov}
\affiliation{Advanced Light Source, Lawrence Berkeley National Laboratory, Berkeley, California 94305, USA}
\author{Hsin Lin}
\author{Arun Bansil}
\affiliation{Department of Physics, Northeastern University, Boston, MA 02115, USA}
\author{Yew San Hor}
\author{Robert J. Cava}
\affiliation{Department of Chemistry, Princeton University, Princeton, NJ 08544, USA}
\author{Liang Fu}
\affiliation{Department of Physics, Harvard University, Cambridge, MA 02138}
\author{M. Zahid Hasan}
\affiliation{Department of Physics, Joseph Henry Laboratories, Princeton University, Princeton, NJ 08544, USA}


\begin{abstract}

The \textbf{Bi$_2$Se$_3$} class of topological insulators has recently been shown to undergo a superconducting transition upon hole or electron doping (Cu$_x$-Bi$_2$Se$_3$ with T$_C$=3.8$^o$K and Pd$_x$-Bi$_2$Te$_3$ with T$_C$=5$^o$K), raising the possibilities that these are the first known ``topological superconductors" or realizes a superconducting state that can be potentially used as Majorana platforms (L.A. Wray \textit{et.al.}, Nature Phys. \textbf{6}, 855-859 (2010)). We use angle resolved photoemission spectroscopy to examine the full details of the spin-orbital groundstates of these materials including \textbf{Bi$_2$Te$_3$}, observing that the spin-momentum locked topological surface states remain well defined and non-degenerate with respect to bulk electronic states at the Fermi level in the optimally doped superconductor and obtaining their experimental Fermi energies. The implications of this unconventional surface (that undergoes superconducting at lower temperatures) topology are discussed, and we also explore the possibility of realizing the same topology in superconducting variants of Bi$_2$Te$_3$ (with T$_C$ $\sim$ 5$^o$K). Characteristics of the experimentally measured three dimensional bulk states are examined in detail for these materials with respect to the superconducting state and topological properties, showing that a single Majorana fermion zero mode is expected to be bound at each superconducting vortex on the surface. Systematic measurements also reveal intriguing renormalization and charge correlation instabilities of the surface-localized electronic modes.

\end{abstract}


\date{\today}

\maketitle

Three dimensional topological insulators (3D-TIs) realize a recently discovered novel state of matter which is \textit{not} reducible to multiple copies of integer quantum Hall states (IQH), in which a topological property of bulk electronic states gives rise to spin-momentum locked two-dimensional Dirac cone surface states (a novel 2DEG) with a host of unusual electronic and spin properties \cite{Intro,TI_RMP,hydAtom,TIbasic,DavidNat1,DavidScience,DavidTunable,MatthewNatPhys,ChenBiTe,WrayCuBiSe,WrayFe,XuTl,XuTernary}. Following the axiomatic principle that ``more is different" in many-body systems, this novel surface or 2DEG environment has been the subject of numerous predictions for emergent many-body states and device physics that cannot be readily realized by other material or quantum Hall-like systems \cite{FuSCproximity,FuHexagonal,FuNew,QiTopoSC,MudrySurfSC,NagaosaUnconventional,PairStr,AshvinChiral,ImpurityStates,FerroSplitting, KitaevClass, SchnyderClass}. Shortly following the first experimental discovery of a three dimensional topological insulator, it was proposed that the introduction of superconductivity at a topological insulator (TI) surface can under the right conditions give rise to braidable (non-Abelian) quasiparticles which might be applied in quantum information science. This proposal has drawn a great deal of attention to the recent observation of superconducting states in doped topological insulators Cu$_x$Bi$_2$Se$_3$ (T$_C$=3.8$^o$K) and Pd$_x$Bi$_2$Te$_3$ (T$_C$=5.5$^o$K) \cite{HorSC,HorAllSC,WrayCuBiSe}. In this paper, we explore the systematic details of the critical questions of band topology and electron dynamics features that constrain the form of superconductivity.  Our results confirm our recent observations that the band structure of superconducting Cu$_{0.12}$Bi$_2$Se$_3$ retains the well defined surface electron kinematics of topological insulator Bi$_2$Se$_3$ \cite{WrayCuBiSe}, and indicate that the surface state in superconducting variants of Bi$_2$Te$_3$ is also likely to be fully preserved at the Fermi level. This topologically ordered electronic system scenario creates a platform in which superconductivity is expected to host non-Abelian Majorana Fermion vortex modes \cite{MudrySurfSC, NagaosaUnconventional, FuNew, QiTopoSC}. The observed electron kinematics are discussed in the context of possible forms that these vortices may take (surface or bulk) and surprising instabilities revealed by Cu-doping in the topological surface band structure over the range x=0 to 12$\%$. These measurements provide important clues for developing a theory of superconductivity in the strongly spin-orbit coupled electronic systems of highly topical Bi-based topological insulators in general.

\begin{figure}[t]
\includegraphics[width = 8cm]{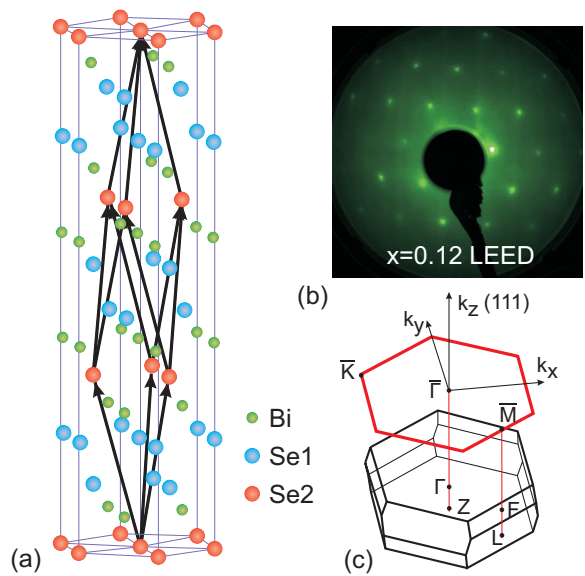}
\caption{(a) The crystal structure of Bi$_2$Se$_3$ is shown. (b) A low energy electron diffraction (LEED) image of superconducting Cu$_{0.12}$Bi$_2$Se$_3$ taken at 200 eV electron energy shows a well ordered surface with no sign of superstructure modulation. (c) The hexagonal surface Brillouin zone of Cu$_x$Bi$_2$Se$_3$ is shown in red above a diagram of the three dimensional bulk Brillouin zone.}
\end{figure}

Early theoretical proposals for achieving non-Abelian surface vortices on a topological insulator followed the concept presented in Ref. \cite{FuSCproximity}, in which superconducting phase coherence is introduced to a topological insulator surface through an interface between the topological insulator and a surface-deposited superconducting material. However, no superconductor has yet been found that can make smooth material contact without a substantial Schottky barrier with the known single crystal topological insulators such as Bi$_2$Se$_3$, Bi$_2$Te$_3$ or Sb$_2$Te$_3$. And furthermore, the presence of a superconducting material overlayer would in itself be an obstacle to experimental probing of vortices on the topological surface. Therefore, the very recent discovery of superconductivity with a significant critical temperature of T$_C$=3.8$^o$K through Meissner effect measurements in Cu-doped Cu$_x$Bi$_2$Se$_3$ \cite{HorSC} was not only a material science breakthrough, but also popularized a new conceptual paradigm in how superconductivity could be incorporated with topological insulator surface states for device purposes. Our measurements of band structure revealing the physical scenario and theoretical expectations presented by superconductivity in Cu$_x$Bi$_2$Se$_3$ are the subject of this paper.

Undoped Bi$_2$Se$_3$ is a topological insulator with a large band gap ($\sim$300 meV) \cite{MatthewNatPhys}, and belongs to a class of materials M$_2$X$_3$ (M=Bi,Sb; X=S,Se,Te) that includes at least two other topological materials, Bi$_2$Te$_3$ and Sb$_2$Te$_3$, with smaller band gaps and more complicated bulk band structures \cite{BiTeSbTe}. They share a rhombohedral crystal structure (see Fig. 1), with a five atom unit cell arranged in quintuple layers and exhibit large thermoelectric power \cite{DiSalvo}. These materials feature almost linearly dispersive spin-polarized conduction electrons on their surfaces, in which electrons behave as massless relativistic particles.

\begin{figure*}[t]
\includegraphics[width = 17.2cm]{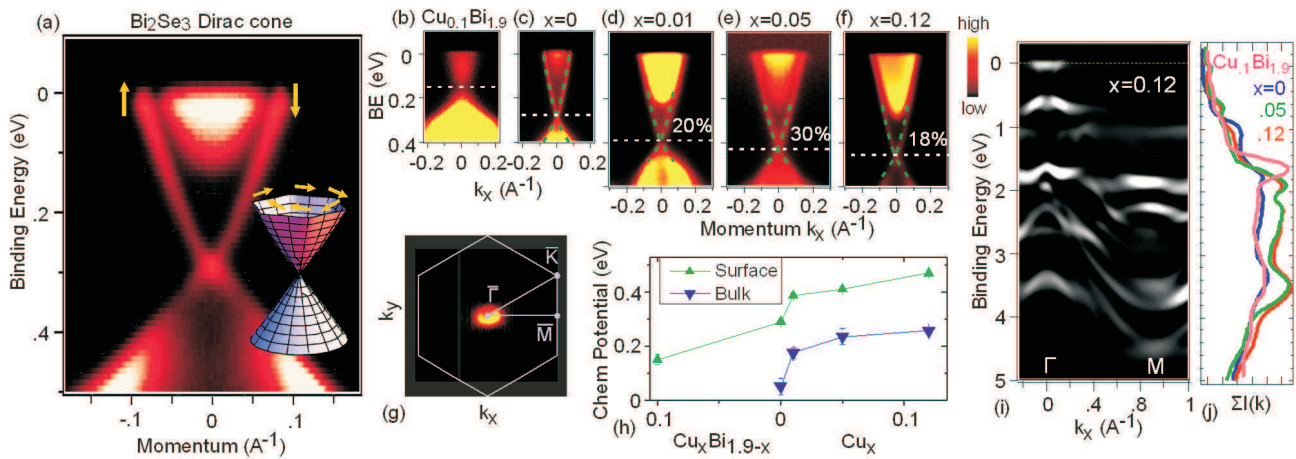}
\caption{\label{fig:dopingCuts}{\bf{Doping and the chemical potential}}: (a) The surface state of Bi$_2$Se$_3$ is a spin-textured Dirac cone. Representative ARPES measurements along the $\Gamma$-$\overline{M}$ axis are shown for dopings spanning the range from (b) a topological insulating state to (f) the optimally doped superconductor. Percentage values indicate the amount by which dispersion in the upper Dirac cone is reduced (renormalized) relative to the undoped compound, and the Dirac point binding energy is indicated with a dashed white line. (g) A low resolution Fermi surface measurement over the full first Brillouin zone is shown for the superconducting compound, performed at high photon energy. (h) Binding energies of the surface Dirac point and bulk conduction band minimum are plotted as a function of doping. (i) A second derivative image of valence bands in the optimal superconductor. (j) Integrated ARPES intensity over the momentum range shown in (i) is plotted for different samples.}
\end{figure*}

Introducing superconductivity through doping of the TI bulk is an optimal approach to bring superconducting phase coherence to the surface of the crystal, but generates a new set of challenges because the topological properties and crystal quality that are desired for device applications can be directly impacted by the dopant. We have found that copper doping leads to increased variability in the crystal quality, and care is needed to grow and identify single-domain crystals (see LEED in Fig. 2(b)) with sharp quasiparticles. Of greater concern, numerical predictions suggest that the topological insulator surface states merge quickly with bulk electronic states as the chemical potential moved from the insulating gap, and the doping required to achieve superconductivity (10$\%$) is far too high for well defined surface states to persist at the Fermi level where Cooper pairing occurs. Even though Hall effect characterization suggested that copper ions donated less than one carrier for each Cu atom \cite{HorSC}, extrapolated predictions based on numerics and band structure measurements of undoped Bi$_2$Se$_3$ all suggest that the topological insulator surface states will have merged with bulk electronic states after 10$\%$ copper doping, making the case for novel surface physics highly uncertain.

However, in the case of Cu$_x$Bi$_2$Se$_3$, nature has proven to be surprisingly accommodating. Direct band structure measurements of Cu$_{0.12}$Bi$_2$Se$_3$ show that the surface and bulk states remain separate at the Fermi level (see e.g. Fig. \ref{fig:inPlaneARPES}), in spite of expectations to the contrary. Using angle resolved photoemission spectroscopy (ARPES) to study the material system over a range of Cu doping values, we observe an atypical renormalization effect that bends the surface bands away from the bulk, and extract band parameters that provide a framework for understanding the novel properties that superconductivity may adopt in the sample surface and bulk, related to the spin-orbit coupled topological insulator band structure. Our measurements show that the conduction band of Cu$_{0.12}$Bi$_2$Se$_3$ resembles an ideal massive 3D Dirac cone, which is the default bulk band shape expected for topological insulators, making the theoretical discussion and analytical methodology generalizable to other material systems.

The ARPES measurements presented here were performed at the Advanced Light Source beamlines 10 and 12 using 35.5-48 eV photons and Stanford Synchrotron Radiation Laboratory (7-22eV photons) with better than 15 meV energy resolution and overall angular resolution better than 1$\%$ of the Brillouin zone (BZ). Samples were cleaved and measured at 15$^o$K, in a vacuum maintained below 8$\times$10$^{-11}$ Torr. Momentum along the $\hat{z}$ axis is determined using an inner potential of 9.75 eV, fine tuned by the photon energy dependence shown in Fig. \ref{fig:zAxisCuts} and consistent with previous photoemission investigations of undoped Bi$_2$Se$_3$ \cite{MatthewNatPhys}. Large single crystals of Cu$_x$Bi$_2$Se$_3$ were grown using methods described in Ref. \cite{HorSC}. Introducing x=0.12 copper doping for optimal superconductivity was found to shift the z-axis lattice parameter by only 1.5$\%$ while leaving the in-plane lattice parameters and long-range crystalline order intact. Surface and bulk state band calculations were performed using the LAPW method implemented in the WIEN2K package \cite{wien2k}, and accurately reproduce the key elements of topological order in the Bi$_2$Se$_3$ system. Details of the calculation are identical to those described in Ref. \cite{MatthewNatPhys}.

This paper is divided into three sections. In the first, we explore the unconventional Cu-doping behavior and instabilities of electrons near the surface, which are critical to preserving the non-degenerate topological band structure. In the second part, we examine the topologically ordered bulk band structure and determine parameters for a material specific model of the superconducting state. In the third part, we comprehensively analyze these results to evaluate how the interplay of topological order and superconductivity are expected to give rise to novel physical properties.

\section{Bulk-doping a topological insulator}

The effect of carrier doping on the surface of a topological insulator is not easy to predict, as it would be for a generic, topologically trivial band insulator. The fact that surface states span a bulk band gap leads to an excess of conducting charge carriers on the surface, causing charge density to be non-uniform along the z-axis. A non-uniform charge density leads to many-body Coulomb interactions that are not readily incorporated in the electronic energies predicted by density functional theory (implemented here in the generalized gradient approximation, DFT-GGA). Furthermore, band energies (and hence carrier density) in the surface state are highly sensitive to the z-axis potential gradient at the surface because of the Rashba effect, adding another level of complexity.

As a result of these and perhaps other effects, we see that addition of copper results in non-linear electron doping and highly unusual changes to the Bi$_2$Se$_3$ surface state kinetics. Because of these properties, copper intercalation techniques may achieve application as an experimental tool to manipulate the surface states in bismuth-based topological insulators such as Bi$_2$Se$_3$ and Bi$_2$Te$_3$, and the effects of copper doping are worthwhile to study in detail. In particular, the reduction in surface state Fermi velocity achieved by copper doping has the effect of increasing the energy/momentum separation between bulk and surface conduction bands, stabilizing the topological surface properties as outlined later in the paper. Based on this, one may also speculate that the addition of copper or a less perturbing neutral-valence dopant could provide a mechanism to restore a topological surface state in material systems for which topological properties have been lost due to overlap between the bulk and surface bands.

\begin{figure}[t]
\includegraphics[width = 8.5cm]{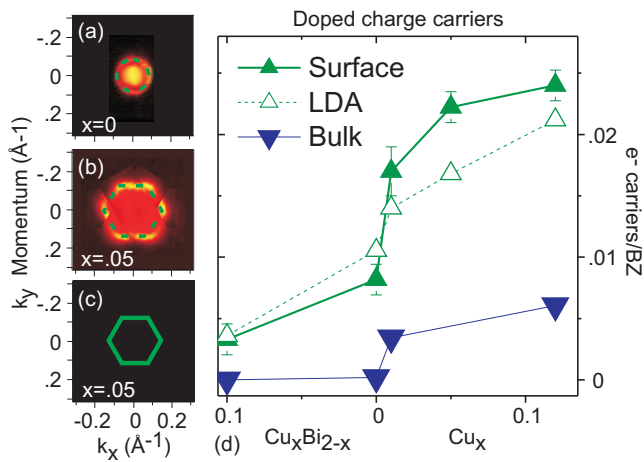}
\caption{\label{fig:FSlutCount}{\bf{Doping towards superconductivity leads to critical changes on the surface electronic and spin environment}}: (a-c), Symmetrized surface state Fermi surfaces are displayed together with (bottom) a DFT-GGA prediction based on the Dirac point binding energy at x=0.05 copper doping. Non-zero intensity inside the panel (b) Fermi surface results from bulk electrons screening the surface state, as described in Fig. 6. (d) The total number of charge carriers in the bulk and at the surface is calculated from the Luttinger count ($\frac{FS\,area}{BZ\,area}$, $\times$2 for the doubly degenerate bulk band). DFT-GGA predictions show the carrier density obtained by aligning DFT band structure with the experimentally determined binding energy of the Dirac point.}
\end{figure}

To gain a systematic understanding of the highly unusual doping effect that copper has on the system, we present data on several stoichiometric crystals with copper added (Cu$_x$Bi$_2$Se$_3$, x=0, 0.01, 0.05, 0.12) and contrast these doping levels with a case in which copper is substituted for bismuth as Cu$_{0.1}$Bi$_{1.9}$Se$_3$. Figure \ref{fig:dopingCuts}(b-f) shows measurements from the data sets that were used to calculate the surface state conduction carrier density (Luttinger count) for Fig. \ref{fig:FSlutCount}(d). These measurements were performed with high photon energies ($>$20 eV) and are used to understand qualitative material properties such as the approximate surface state band contour; however we have found that lower photon energies are necessary to resolve finer details of the bulk band structure due to a screening effect discussed with respect to Fig. \ref{fig:TIarpes}.

Copper addition gradually enlarges the Fermi surface through electron doping and, surprisingly, brings about a strong reduction in surface electron velocities through reshaping of the surface state conduction band. The slope of the surface state band represents particle velocity, and is reduced by an increasing amount as the Cu concentration is raised to 5$\%$, by which point electron velocities near the Dirac point are 30$\%$ slower than at the Dirac point of undoped Bi$_2$Se$_3$. The surface electron velocities recover slightly at superconducting doping but are still approximately 20$\%$ slower than in the undoped compound. Higher binding energy valence band structure evolves through a rigid shift as doping is changed (Fig. \ref{fig:dopingCuts}(j)), suggesting that with the exception of adding electrons, copper does not significantly alter the bulk electronic properties or perturb the more localized (tightly bound) electronic states. The energy spacing of peak features at binding energies greater than 1 eV in the integrated density of states are essentially independent of doping, implying that renormalization of the surface state conduction band is not due to a change in the bulk material Hamiltonian such as the strength of spin orbit coupling.

As the surface state Fermi surface enlarges due to added carriers it also becomes hexagonally anisotropic, as outlined in Fig. \ref{fig:FSlutCount}(a-c). Such hexagonal-like deformation makes the surface susceptible to spin-fluctuation or magnetic instabilities \cite{FuHexagonal}. Tunability of the surface state kinetics and anisotropy is important for control of unconventional ordered states that may appear uniquely in topologically ordered materials. Carrier density in the surface state is much greater than density in the bulk, as estimated by the Luttinger count (Fig. \ref{fig:FSlutCount}(d)), suggesting that the surface may carry a screened negative charge within the normal state of the superconductor. Attempting to force the creation of Cu$_{Bi}$ replacement defects by adding less bismuth results in very weak hole doping for Cu$_{0.1}$Bi$_{1.9}$Se$_3$ (Fig. \ref{fig:dopingCuts}(h)), raising the bulk conduction band entirely above the Fermi level so that the material is a traditional topological insulator. These results establish that copper atoms can add holes or electrons depending on their net preference for occupying different kinds of site in the Bi$_2$Se$_3$ lattice. Copper atoms intercalated between van der Waals-like bonded selenium planes (Cu$_{int}$) in the crystal are thought to be single electron donors, while substitutional defects in which copper replaces bismuth in the lattice (Cu$_{Bi}$) contribute two holes to the system \cite{CuAmphoteric}. The bulk doping levels observed in our data suggest that the relative number of copper atoms occupying these two configurations evolves from a ratio of $\frac{Cu_{int}}{Cu_{Bi}}$=3.5 at x=0.01 to just 2.2 in superconducting x=0.12 crystals.

\begin{figure}[t]
\includegraphics[width = 7cm]{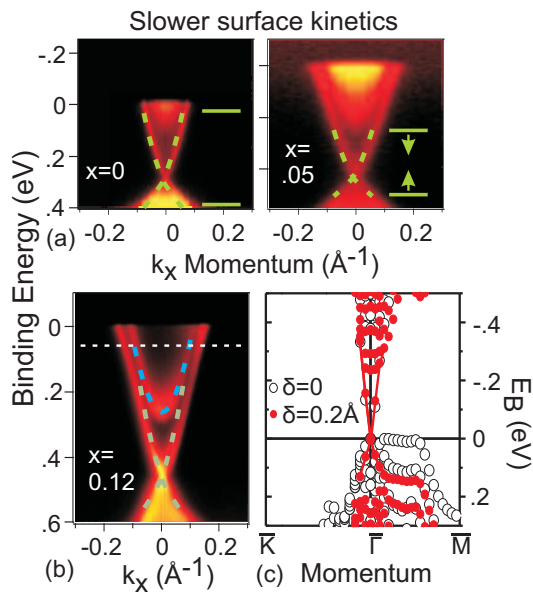}
\caption{\label{fig:slowVel}{\bf{Doping towards superconductivity leads to changes on the surface band (and Fermi) velocities}}: (a) Photoemission measurements at high photon energy (E$>$20 eV) are shown for non-superconducting Cu$_x$Bi$_2$Se$_3$. Particle velocities in the surface state upper Dirac cone are reduced by 30$\%$ after the addition of x=0.05 copper doping. The energy scale of panel (a,right) is offset by a rigid shift to align the Dirac point binding energy with the panel at left. (b) The surface state dispersion from undoped Bi$_2$Se$_3$ is overlaid in green on a photoemission image of Cu$_x$Bi$_2$Se$_3$ at superconducting composition (x=0.12), measured with low energy (9.75 eV) photons to observe (blue) the bulk dispersion. By linear extrapolation, we project that if the surface state velocity were not reduced after Cu doping, the bulk and surface bands would intersect below the Fermi level at the energy labeled with a dashed white line. (c) When a 12-quintuple-layer slab calculation is modified by shifting the outermost (Se) atomic layer 0.2 $\AA$ away from the next (Bi) atomic layer (layer spacing increased from 1.64–1.84 $\AA$), dispersion in the upper Dirac cone increases by 16$\%$. This calculation demonstrates that the surface state dispersion is very sensitive to minor changes in the surface chemical environment.}
\end{figure}

\begin{figure}[t]
\includegraphics[width = 8.7cm]{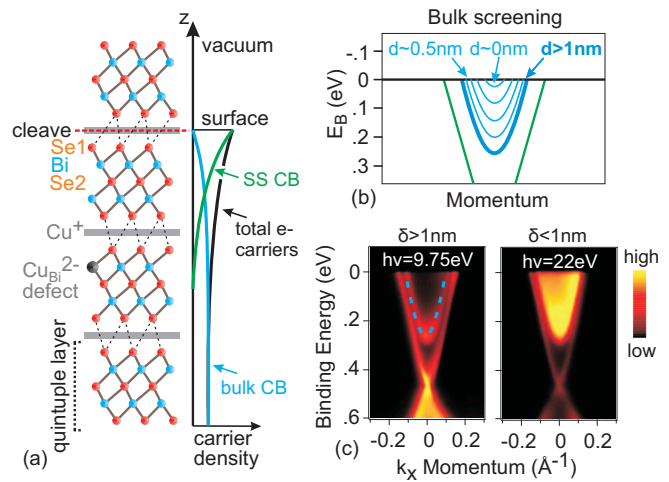}
\caption{\label{fig:TIarpes}{\bf{Bulk electrons screen the surface charge}}: (a) The crystal structure of Cu$_x$Bi$_2$Se$_3$, viewed parallel to the (111) cleavage plane. Copper can donate electrons when intercalated between Se layers, or may accept electrons if it substitutes for Bi. The bulk and surface state electron density distributions are sketched as discussed in the text. (b) The large surface state carrier density is screened by the bulk band, causing ``bulk" states to be hole doped (occur at smaller binding energy) in close proximity to the surface. (c) ARPES measurements with (left) large and (right) sub-nanometer penetration depth are compared.}
\end{figure}

Reducing the slope of the surface state increases separation between the bulk and surface conduction bands, a characteristic that preserves stability of the topological order. In Fig. \ref{fig:slowVel}(b), the undoped Bi$_2$Se$_3$ surface dispersion is overlaid on the band structure of x=0.12 Cu$_x$Bi$_2$Se$_3$, demonstrating that with no reduction in the undoped band velocity, it is not expected that the surface and bulk bands will remain separate. The calculation in Fig. \ref{fig:slowVel}(c) suggests one possible mechanism by which the surface band velocity could be reduced. Increasing the spacing between the outermost two atomic layers by just 0.2 $\AA$ from the bulk equilibrium yields a much ``pointier" surface Dirac cone that is in better agreement with the experimentally observed shape, and increases the upper Dirac cone dispersion by 16$\%$ while lowering the energy of states in the lower Dirac cone. One explanation for reduced velocity may therefore be that the crystal surface of undoped Bi$_2$Se$_3$ is more strongly relaxed (lowering the valence band energy), but as more electrons are added to the conduction band it becomes energetically favorable to lower the conduction band energies by reducing the interatomic spacing to a value closer to its bulk equilibrium position. Other possible explanations and contributing factors exist and will require extensive experimental characterization as mechanisms for tuning the surface properties of topological insulators.


The surface of a topological insulator is a complex environment for photoemission measurements due to the vertical-axis anisotropy and strong topological protection of the Dirac-like surface states. Although DFT-based calculations predict only a single surface conduction state, surface sensitive ARPES measurements performed at high photon energy show an additional distortion localized above the bulk conduction band in energy, which appears to occur separately from the topological surface conduction state. When the band is measured with higher bulk sensitivity (Fig. \ref{fig:TIarpes}(c,left)), it can be clearly resolved as a narrow positive-mass paraboloid, however the inside of the paraboloid becomes filled in with intensity when higher photon energies are used (h$\nu$$>$20 eV, see Fig. \ref{fig:TIarpes}(c,right)). Figure \ref{fig:dopingCuts} contains further examples of filled-in bulk band structure.

A likely explanation for this phenomenon relates to the many-body Coulombic interaction between electrons at the surface and in the bulk, which is not accounted for by DFT. As we have observed through the systematic doping measurements summarized in Fig. \ref{fig:FSlutCount}(d), the surface state carrier density of Cu$_x$Bi$_2$Se$_3$ and Cu$_{0.1}$Bi$_{1.9}$Se$_3$ is much greater than the carrier density in the bulk, and holds a negative charge. Theory suggests that surface state penetration into the crystal is approximately given by a ratio of the surface state band velocity and the bulk band gap (v$_{S}$/$\Delta$), which is less than 1 nm for Cu$_x$Bi$_2$Se$_3$ (3eV$\AA$/0.4 eV = 7.5$\AA$ for x=0.12). Given this scenario, the bulk and surface carrier densities (and charge density) are expected to conform to a distribution similar to that drawn in Fig. \ref{fig:TIarpes}(a). We note that details of the absolute charge density distribution may depend on the broader sample environment and boundary conditions.

\begin{figure}[t]
\includegraphics[width = 8cm]{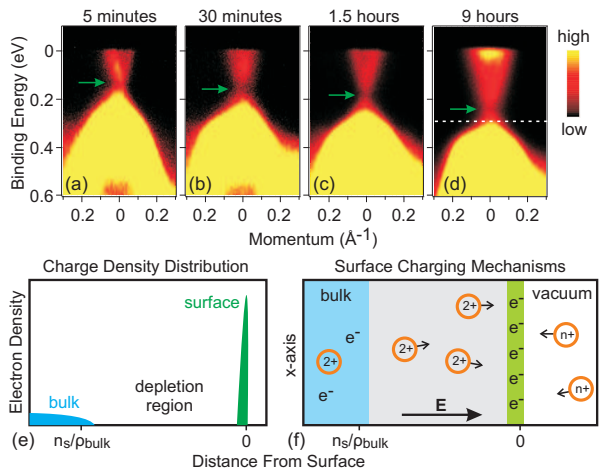}
\caption{\label{fig:timeEvolve}{\bf{Time evolution of Cu-substituted Cu$_{0.1}$Bi$_{1.9}$Se$_3$}}:  Unlike Cu-intercalated Cu$_x$Bi$_2$Se$_3$ The chemical potential in Cu-substituted Cu$_{0.1}$Bi$_{1.9}$Se$_3$ slowly rises over time after the surface is exposed to vacuum. The binding energy of the surface state Dirac point is indicated with an arrow for spectra measured (a) 5 minutes (b) 30 minutes (c) 1.5 hours and (d) 9 hours after cleavage. The undoped Dirac point binding energy is indicated with a dashed white line in panel (d). Alignment of crystalographic axes in (a) differs slightly from other panels. (e) The charge density distribution expected near the surface of a highly insulating, freshly cleaved Bi$_2$Se$_3$ sample is shown. (f) Two mechanisms that can potentially cause the surface chemical potential to shift upwards over time are the migration of positively charged selenium vacancy defects attracted by the negatively charged surface state, and the adsorption of positive ions from the vacuum (mostly H$^+$).}
\end{figure}

Panel \ref{fig:TIarpes}(b) shows an approximation of hole doped states the bulk electrons may occupy to screen the surface state within 1 nm of the surface. These states constitute a continuum of upwardly shifted ``band images" that fill in the the inside of the bulk dispersion due to (potentially quite complex) Coulomb interaction with the surface state. Because the screening effect is highly localized at the surface, these hole doped band images have broken translational symmetry along the z-axis, and are generally somewhat visible at arbitrary k$_z$ values (e.g. see Fermi surface maps in Fig. \ref{fig:FSlutCount}(a-c)). In a classical picture based on the Poisson equation, it is expected that screening will occur over less than 0.5 nm \cite{fisherBending}, but this is clearly a non-physical model for detailed predictions on such a short length scale because of the discrete nature of the lattice, delocalization of the surface state and numerous related concerns. Qualitatively, reduction in the screening length as carrier density increases is expected to strengthen hybridization between bulk and surface electronic states, and may be partly responsible for the reduced surface state band velocities we observe.

Topological insulator surfaces are often unstable over a long period in vacuum, becoming increasingly electron doped over a period of minutes or hours (see e.g. Ref.\cite{DavidTunable,BiTeSbTe}). Of the Cu-doped samples studied, only the surface of Cu$_{0.1}$Bi$_{1.9}$Se$_3$ was observed to change over a 24 hour time period, with a gradually increasing chemical potential as shown in Fig. \ref{fig:timeEvolve}(a-d). The chemical potential rises quickly in the first hour and a half, with approximately 0.002 electrons added per surface Brillouin zone, however it takes another 7.5 hours to add 0.002 more electrons, suggesting that the effect saturates or will only occur for highly insulating samples. Theories suggest that this time dependence is unrelated to the fundamental topological order of topological insulator band structure\cite{TIbasic}, but it nonetheless constitutes a noteworthy behavior of the known topological insulator compounds. Recent experimental studies in Ref. \cite{WrayFe} have shown that strongly perturbing the surface energetics of a topological insulator results in the formation of new surface bands and Dirac points, in configurations that respect the topological insulator order.

Analyses have shown that the rising surface chemical potential is probably due to gradual accumulation on the surface of positively ionizing gas molecules \cite{YulinFe}, and it is well established that depositing positive ions on the crystal surface will donate negative charge carriers \cite{DavidTunable,BiTeSbTe,WrayFe}. The data presented here suggest one additional instability that may lead to a similar effect, namely that the negative charge we have observed in the surface state will attract positively charged selenium vacancy defects, causing the surface to gradually become more strongly electron doped as summarized in Fig. \ref{fig:timeEvolve}(e-f). Such an effect would weaken greatly as the surface chemical potential approaches the bulk band. The electric field from the surface is screened over a distance similar to the ratio of surface and bulk charge densities, which may be tens of nanometers or longer for highly insulating samples \cite{fisherBending}.

These systematic doping studies show an unexpectedly large spectrum of surface electronic behavior, from hexagonal warping of the Fermi surface to strong renormalization of band velocities, all of which occurs while the chemical potential changes by an unexpectedly small amount. These features highlight the strong sensitivity of the surface state kinetics to intrinsic properties such as the relaxation of lattice spacing near the cleaved crystal surface and complicated interactions between the surface and bulk electrons.


\section{Bulk Electron Kinetics and Superconductivity}

\begin{figure}[t]
\includegraphics[width = 8.7cm]{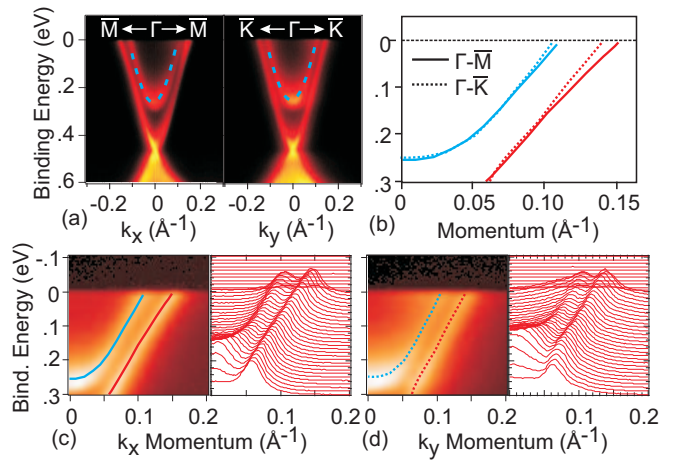}
\caption{\label{fig:inPlaneARPES}{\bf{Band topology at superconducting composition}}: (a) Momentum dependence of the bulk and surface conduction bands in superconducting Cu$_{0.12}$Bi$_2$Se$_3$ is measured through the 3D Brillouin zone center with low energy (9.75eV) photons for enhanced bulk sensitivity. (b) Dispersion along the $\Gamma$-$\overline{M}$ and $\Gamma$-$\overline{K}$ directions is traced from the blown-up measurements shown in panels (c-d), revealing significant anisotropy in the surface state dispersion.}
\end{figure}

The compounds Cu$_x$Bi$_2$Se$_3$ and Pd$_x$Bi$_2$Te$_3$ are the first representatives of a novel class of superconducting materials with strong spin-orbit coupling. Unlike conventional superconductors or noncentrosymmetric superconductors with spin-split bands, these materials have an inversion symmetric crystal structure with doubly degenerate bulk bands. In this case, spin-orbit coupling leads to strong interband mixing that cannot be captured by any single-band model, and is the origin of the topological insulator state. As a starting point for studying bulk superconductivity in Cu$_x$Bi$_2$Se$_3$, the spin-orbit coupled character of its band structure needs to be examined in detail.


\begin{figure}[t]
\includegraphics[width = 8cm]{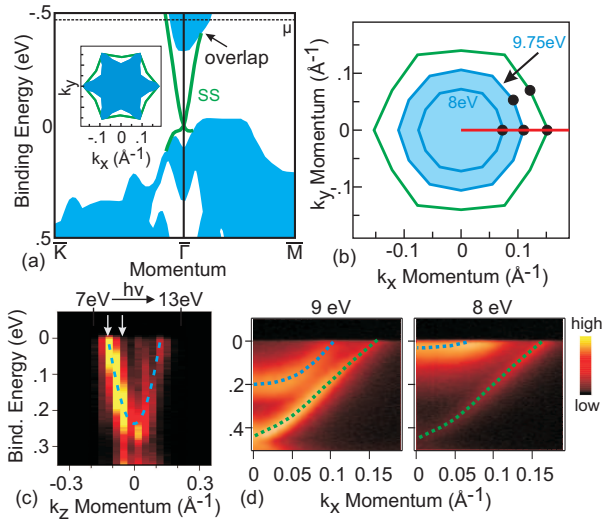}
\caption{\label{fig:zAxisCuts}{\bf{Band topology along the $\overline{\Gamma}$-$\overline{M}$ axis}}: (a) DFT-GGA calculations predict that the surface and bulk conduction band structure will overlap first (at highest binding energy) along the $\overline{\Gamma}$-$\overline{M}$ axis. An inset shows the predicted overlapping Fermi surface at the experimentally derived chemical potential, with the k$_z$-projected bulk band shaded in blue. (b) The experimentally determined Fermi surface contours are shown, with black dots marking specific Fermi momenta identified from panel (d) and Fig. 7. (c) Dispersion along the z-axis is examined by varying the incident photon energy. (c) Dispersion along $\overline{\Gamma}$-$\overline{M}$ is shown for momenta spanning the bulk k$_z$ dispersion to establish that there is no degeneracy between the bulk and surface state electrons at k$_z\neq$0.}
\end{figure}

Doping into the superconducting regime (x=0.12) raises the Fermi energy to 0.25 eV above the bulk conduction band minimum, placing the Fermi level in a highly linear regime of bulk band dispersion (Fig. \ref{fig:inPlaneARPES}). Therefore, Cooper pairing in the superconducting state is characterized by the properties of these linearly dispersive electronic states close to the chemical potential. Bulk band structure with a ``relativistic" lineshape that transitions from parabolic to linear as momentum increases is a direct consequence of band inversion required for the topological insulating state, and has also been experimentally observed in topological insulator Bi$_{1-x}$Sb$_x$ alloys \cite{DavidNat1}. In Bi$_2$Se$_3$ band inversion occurs at the $\Gamma$-point (k$_x$=k$_y$=k$_z$=0) in the center of the bulk conduction band leading to the expectation of a spin-orbit induced Dirac-like bulk band. In our experiments this is indeed the case - the bottom of the conduction band for electrons in the bulk is found at the three dimensional $\Gamma$-point, which also lies inside of the upper surface state (SS) Dirac cone without touching it. Bulk Fermi momenta of 0.110$\pm$3 $\AA^{-1}$ and 0.106$\pm$3 $\AA^{-1}$ are observed along the $\overline{\Gamma}$-$\overline{M}$ and $\overline{\Gamma}$-$\overline{K}$ directions respectively. Carefully tracing the band (Fig. \ref{fig:inPlaneARPES}(b-d)) reveals a Fermi velocity of 3.5 eV$\times\AA$ along $\overline{\Gamma}$-$\overline{M}$ and 4.1 eV$\times\AA$ along $\overline{\Gamma}$-$\overline{K}$, estimated within 50 meV of the Fermi level. The gap between bulk valence and conduction bands appears to be unchanged upon copper doping, consistent with the observation from Section I that the spin orbit coupling strength and bulk Hamiltonian appear to be unchanged. These experimentally determined bulk and surface Fermi surfaces are convex, while the numerically predicted Fermi surfaces are concave (Fig. \ref{fig:zAxisCuts}(a-b)), emphasizing the critical importance of experimental investigations to understand the band structure and electronic states of topological insulators. The numerics can be approximately reconciled with experimental measurements by multiplying ("renormalizing") DFT-derived band energies by a factor of 2 above the bulk conduction band minimum.

The kinetic behavior of electrons with momentum perpendicular to the cleaved surface is examined in Fig. \ref{fig:zAxisCuts}(b-d) by varying incident photon energy. The z-axis Fermi momentum observed in Fig. \ref{fig:zAxisCuts}(c) is 0.12$\pm$1 $\AA^{-1}$, suggesting that the bulk electron kinetics are three dimensionally isotropic with only slightly reduced velocities in the out-of-plane direction. The essential kinetics of a positive-mass band with less than 1$\%$ carrier occupancy as is the case here can generally be evaluated from measurements that pass through the band minimum along the three high-symmetry directions (k$_x$, k$_y$ and k$_z$). However, band structure calculations suggest that bulk dispersion along the $\overline{\Gamma}$-$\overline{M}$ axis will intersect the surface state band before any other location (Fig. \ref{fig:zAxisCuts}(a)), and may be less-than-parabolic for k$_z$$\neq$0, prompting a closer investigation. Band structure measurements shown in Fig. \ref{fig:zAxisCuts}(d) examine dispersion along the $\overline{\Gamma}$-$\overline{M}$ axis at out-of-plane momenta spanning the bulk k$_z$ dispersion, confirming that the separation between bulk and surface bands increases monotonically as momentum is increased along the z-axis. Exploring the full momentum space reveals that the surface and bulk electronic states at the Fermi level are well separated by 0.04 $\AA^{-1}$ in momentum and an energy spacing of about $\Delta_E$=130 meV, with closest proximity in the k$_z$=0 plane.

\begin{figure}[t]
\includegraphics[width = 7.5cm]{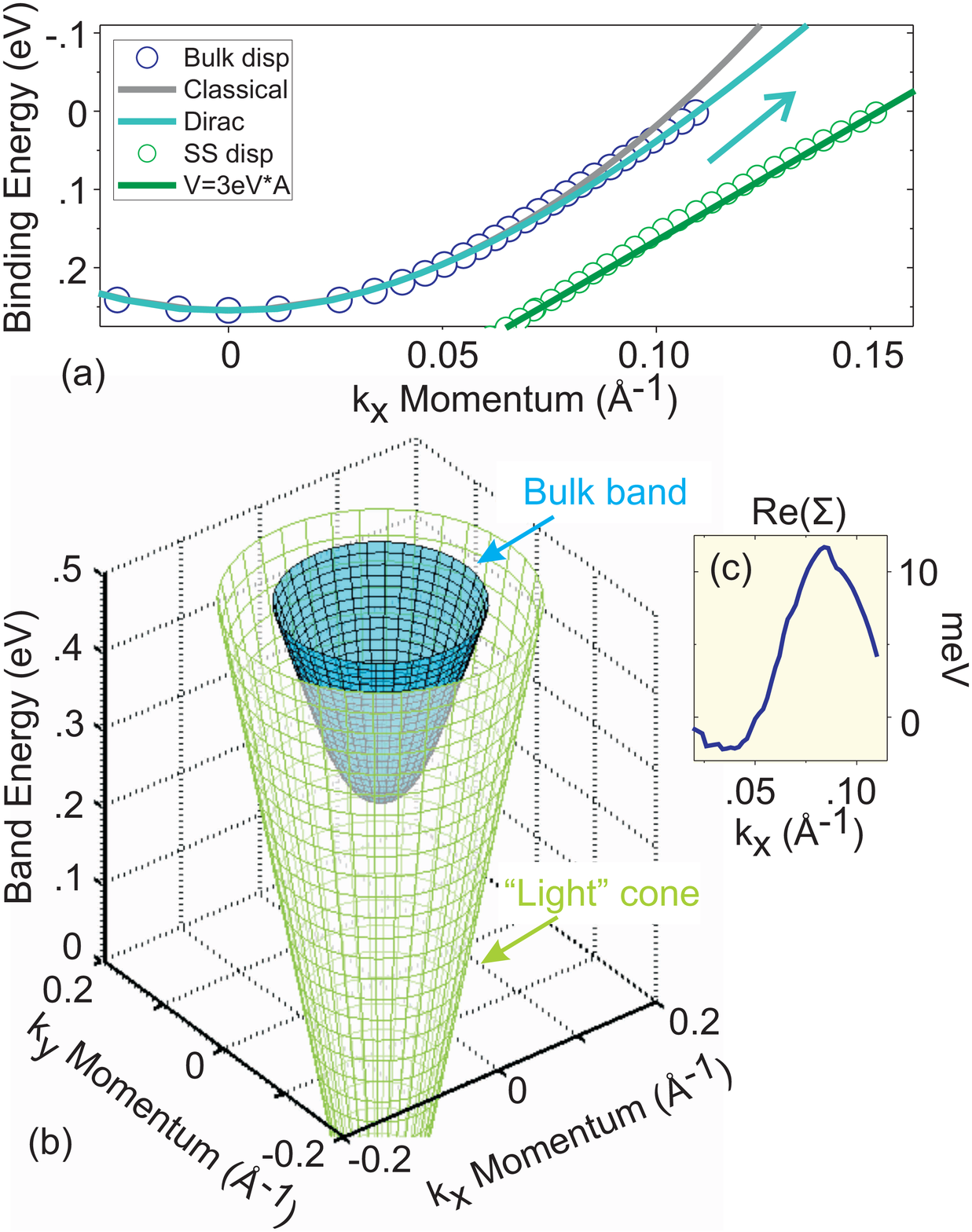}
\caption{\label{fig:DiracFit}{\bf{Massive 3D Dirac fermions in Cu$_x$Bi$_2$Se$_3$}}: (a) Energy-momentum dispersion of bulk electrons through the 3D BZ center is compared with Dirac-like (v=6 eV$\cdot$$\AA$) and classical (parabolic) fits with the same effective mass of 0.155M$_e$. Surface-state (SS) dispersion is plotted with a massless linear fit. (b, blue) The bulk band dispersion approximated from photoemission data is plotted at k$_z$=0, surrounded by (green) a massless ``light-like" Dirac dispersion with identical Dirac velocity (v$_c$), illustrating that bulk electron kinetics are in the relativistic Dirac regime. (c) A small panel shows the difference between measured bulk band energy and the Dirac-like fit curve in panel (a).}
\end{figure}


A minimal two-band model for the band structure of Cu$_x$Bi$_2$Se$_3$, which correctly takes into account of
spin-orbit coupling and crystal symmetries, has been proposed recently\cite{FuNew}. The $k \cdot p$
Hamiltonian near $\Gamma$ in the 3D Brillouin zone is given by
\begin{eqnarray}
H_{0}(\mathbf{k}) = m \sigma_x +v(k_{x} \sigma_z   s_y - k_{y} \sigma_z  s_x)+
v_{z}k_{z}\sigma_y,  \label{H0}
\end{eqnarray}
Here $\sigma$ matrices are Pauli matrices acting on a two-band basis. The eigenvalues of $\sigma_x =1(-1)$ label the conduction (valence) band at $\bk =0$, which is mainly an even (odd) linear superposition of two $p_z$ orbitals on the top and bottom Se-Bi bilayer within the five-layer unit cell. The parameter $s$ represents Pauli matrices labeling electron spin, and $k_z$ is along the [111] direction perpendicular to the cleavage plane of the rhombohedral crystal. As shown in (\ref{H0}), the spin-orbit coupling term is off-diagonal in band basis. The form of $H_0$ resembles
the Dirac Hamiltonian for massive electron and position in $3+1$-dimensional QED. This becomes transparent when
we introduce a set of $4 \times 4$ Gamma matrices: $\Gamma_0 \equiv \sigma_x, \; \Gamma_1 \equiv \sigma_z s_y, \;
\Gamma_2 \equiv - \sigma_z s_x, \; \Gamma_3 \equiv \sigma_y$, so that $H_0= m \Gamma_0 + v (k_x \Gamma_1 + k_y \Gamma_2)
+ v_z k_z \Gamma_3$. We note that the velocity of the massive Dirac fermion is anisotropic due to lack of rotational symmetry in the crystal.

We now fit the band dispersion measured by ARPES with that of the model Hamiltonian $H_0$:
\begin{eqnarray}
E_{\pm}= \pm \sqrt{m^2 + v^2 k_x^2 + v^2 k_y^2 + v_z^2 k_z^2}.
\end{eqnarray}

The curves in Fig. \ref{fig:DiracFit}(a) demonstrate that the conduction band can be closely described using a Dirac mass of m=0.155 M$_e$ (electron masses) and Dirac velocity of v=6 eV$\cdot$$\AA$. Contrasting this fit with a classical (parabolic) dispersion of identical mass reveals that electron kinetics begin to enter the linear relativistic regime within $\sim$0.1 eV of the Fermi level Fig. \ref{fig:DiracFit}(a-b). A slight bend in the dispersion centered near 90 meV binding energy (Fig. \ref{fig:DiracFit}(c)) may suggest strong electron-boson interactions in the system also ubiquitously observed in other superconducting materials \cite{SCproperties} or a shift in the balance of orbital and spin-orbit terms defining the electron kinetics. This model effectively addresses symmetry mixing from spin-orbit coupling and the qualitative shape of the conduction and valence bands, but does not capture fine details of the valence band dispersion. Experimental and numerical studies have shown that the bulk valence band of topological insulators in the M$_2$X$_3$ family is warped \cite{MatthewNatPhys,BiTeSbTe}, with a local minimum (rather than maximum) at the Brillouin zone center due to orbital-kinetic terms dominating over spin orbit interactions to reverse the sign of electron velocity for a small range of momenta. Spin-orbit symmetry mixing captured by Equations 1-2 is very weak in the immediate vicinity of the $\Gamma$-point, allowing for significant deviations of this sort, but quickly becomes the dominant kinetic term at larger momenta where Cooper pairing takes place.


The observed spin-orbit band structure kinematics define several key parameters that give insight into aspects of the superconducting wavefunction without the need for a detailed model. The average Fermi velocity observed from the measurements shown in Fig. \ref{fig:inPlaneARPES}(b-d) suggests that the superconducting coherence length is approximately $2000\AA$ ($\xi_0$$\sim0.2\times\hbar v_F/K_BT_C=0.2\times3.8eV\AA/(K_B\times3.8^oK)=2000\AA$), assuming minimal scattering and a superconducting gap related to T$_C$ by a BCS-like perturbative Cooper pairing mechanism. This value is about 1-2 orders of magnitude greater than the correlation lengths estimated from band structure of the cuprates ($\xi_0$$\sim100-200\AA$) or cobaltates ($\xi_0$$\sim200\AA$) \cite{SCproperties,SCDong}. Correlation length and carrier mobility set the phase ordering temperature scale, which can thus be estimated to be T$_\theta$ $\sim$ 60,000$^o$K (T$_\theta$= $\xi_0\times\hbar^2n_e/2m^*$= 2000$\AA\times\hbar^2\times10^{20} cm^{-3}/(2\times.155 Me)$ = 60,000$^o$K), four orders of magnitude larger than the superconducting critical temperature of 3.8$^o$K. These properties of large coherence length and high phase ordering temperature suggest that the superconducting phase transition and bulk superconducting gap formation can be understood within a mean field picture, although to describe the spin-orbit symmetries and collective properties of the topologically ordered superconducting electrons one needs a material specific model as we will discuss further in Section III. Based on mean field considerations, we expect the superconducting gap to be about $\sim$0.6 meV (3.5$\times$K$_B$T$_C$/2=0.6 meV), much smaller than the state-of-the-art resolution of ARPES.

\begin{figure}[t]
\includegraphics[width = 8.7cm]{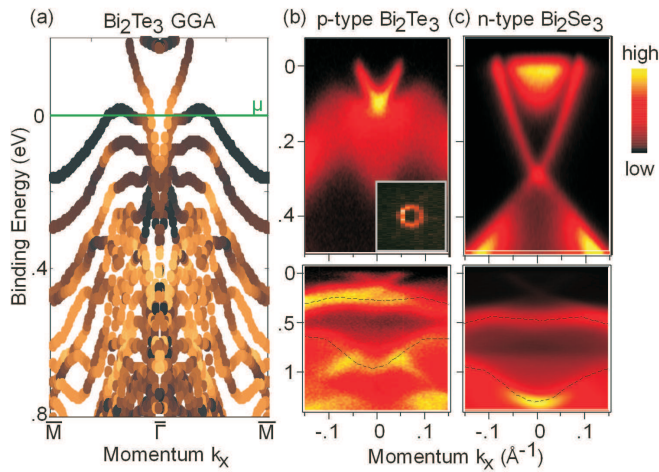}
\caption{\label{fig:BiTeSurfID}{\bf{Band topology in superconducting composition of Bi$_2$Te$_3$}}: (a) The band structure of Bi$_2$Te$_3$ is predicted by the generalized gradient approximation (GGA) for a six quintuple-layer slab, with lighter colors marking states with greater occupancy of the outermost quintuple layer. (b-c) The band structures of Bi$_2$Te$_3$ and Bi$_2$Se$_3$ measured by ARPES with high photon energies ($>$20eV) are shown side by side, and an inset displays the Bi$_2$Te$_3$ Fermi surface in a $\pm$0.2$\AA^-1$ window, revealing only the surface state. Panels at the bottom show band structure at higher binding energy, with dashed lines highlighting a pattern of surface resonance that is common to topological insulators in the M$_2$X$_3$ series. The spectrum in panel (b,top) was measured immediately after cleavage, before any significant observed shift in the surface chemical potential.}
\end{figure}

The spin textured topological surface state is key to many of the most interesting properties that can emerge from the interplay of topological order and broken symmetry. Therefore for any superconducting doped topological insulator it is critical to experimentally identify if after doping the surface state is fully preserved at the Fermi level where Cooper pairing occurs, as our measurements shown in Fig. \ref{fig:inPlaneARPES} and Fig. \ref{fig:zAxisCuts} have established to be the case for Cu$_{0.12}$Bi$_2$Se$_3$. As identified earlier in this Section, the separation of bulk and surface electronic states is 0.04 $\AA^{-1}$ in momentum and $\Delta_E$=130 meV in energy, defining the scale on which perturbations can be applied for device engineering without compromising the capability to form collective electronic states that are localized entirely at the surface. For the surface state to be well defined, it must be fully separate (non-degenerate) from the k$_z$-projected bulk band continuum at the Fermi level. A scenarios in which this condition is not met is demonstrated by the numerically predicted Fermi surface in Fig. \ref{fig:zAxisCuts}(a), and the important role played by unusual surface instabilities to achieve a non-degenerate surface state in the Cu$_x$Bi$_2$Se$_3$ material matrix has been discussed in Section I.

Now that superconducting states with large transition temperatures have been identified in chemically (Pd$_x$Bi$_2$Te$_3$, T$_C$=5.5 $^o$K) \cite{HorAllSC} and mechanically strained (T$_C$$\sim$3 $^o$K) \cite{ZhangBiTePressure} versions of the closely related compound Bi$_2$Te$_3$, it is desirable to perform the same sort of band structure analyses for that system. Numerical simulations such as that shown in Fig. \ref{fig:BiTeSurfID}(a) suggest that the surface state upper Dirac cone at the superconducting chemical potential will be well defined and found at smaller momenta than the bulk conduction states. However, photoemission measurements (Fig. \ref{fig:BiTeSurfID}(b)) have yet to reveal any sign of the bulk conduction bands at the Fermi level, in spite of a significant p-type Hall carrier concentration of 1$\times$10$^{19}$cm$^{-3}$. The carrier density of p-type as-grown Bi$_2$Te$_3$, shown in Fig. \ref{fig:BiTeSurfID}(b), is thought to be similar to superconducting variants of Pd$_x$Bi$_2$Te$_3$, which cannot yet be grown with a high degree of crystalline homogeneity. A region of photoemission intensity from 0.1 to 0.3 eV binding energy merges with the lower Dirac cone, and is blurred in a pattern that resembles the surface-screened bulk conduction band of Bi$_2$Se$_3$, but has no visible intersection with the Fermi level.

As with undoped Bi$_2$Se$_3$, the measured band structure of Bi$_2$Te$_3$ is characterized by larger group velocities than are predicted by first principles calculations. The measured $\overline{\Gamma}$-$\overline{M}$ Fermi velocity of the surface state is 1.9 eV$\AA$, as compared to 2.3 eV$\AA$ in the slab calculation. Surface resonance features at larger binding energies such as the diamond shape traced in the bottom panels of Fig. \ref{fig:BiTeSurfID}(b-c) are shortened on the energy axis by a factor of $\sim$2 in the Fig. \ref{fig:BiTeSurfID}(a) calculation, similar to the observed renormalization of the Cu$_x$Bi$_2$Se$_3$ conduction band. With no definite experimental observation of the bulk conduction states, and significant quantitative uncertainty in first principles calculations, it is currently challenging to determine whether or not the bulk and surface states of superconducting Bi$_2$Te$_3$ overlap.

The bulk band structure of topological insulators has previously been studied to only a limited degree because of the screening effect described in Fig. \ref{fig:TIarpes} and the bulk insulating character, which limits experimental investigation of the conduction band. Due to the low photon energies used and electron doped chemical potential, the results presented in this section give a clear view of the bulk conduction band kinetics of Bi$_2$Se$_3$, a compound that has been termed a ``hydrogen atom" topological insulator for its simplest-case realization of topological insulator \cite{hydAtom,MatthewNatPhys}. These band parameters, including a nearly isotropic massive Dirac dispersion and the complete separation of surface and bulk conduction states provide the key details of electron kinetics and topology to formulate a theory of the novel superconducting properties of Cu$_{0.12}$Bi$_2$Se$_3$. As we will discuss in Section III, superconductivity at the surface of Cu$_{0.12}$Bi$_2$Se$_3$ cannot be conventional given the observed band properties, and is expected to be the first realization of one of two novel forms of superconductivity characterized by electronic topology.

\section{Implications: Non-Abelian Superconductor or Topological Superconductor?}

\begin{figure}[t]
\includegraphics[width = 7.5cm]{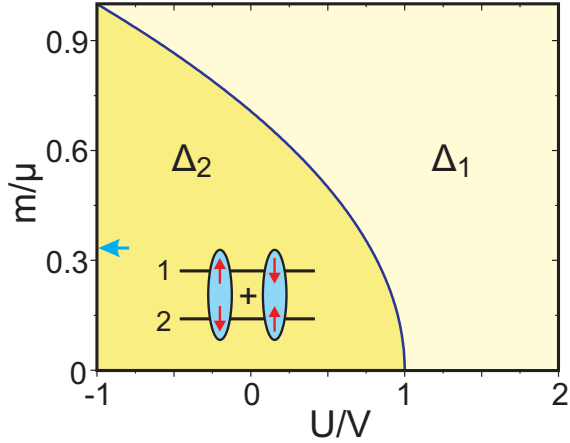}
\caption{\label{fig:FuPhaseDiag}{\bf{Surface Majorana vortices vs. topological superconductor}} The ``m/$\mu$" ratio derived from ARPES parameters is critical to defining whether the ground state will be a ($\Delta_2$-state) topological superconductor with odd parity superconductivity, or ($\Delta_1$-state) will host non-Abelian Majorana Fermion surface vortices with an even parity superconducting wavefunction. Parameters ``U" and ``V" represent matrix elements for intraorbital and interorbital scattering amplitudes from the superconducting pairing interaction, as described in the text and Ref. \cite{FuNew}.}
\end{figure}

\begin{figure}[t]
\includegraphics[width = 7.8cm]{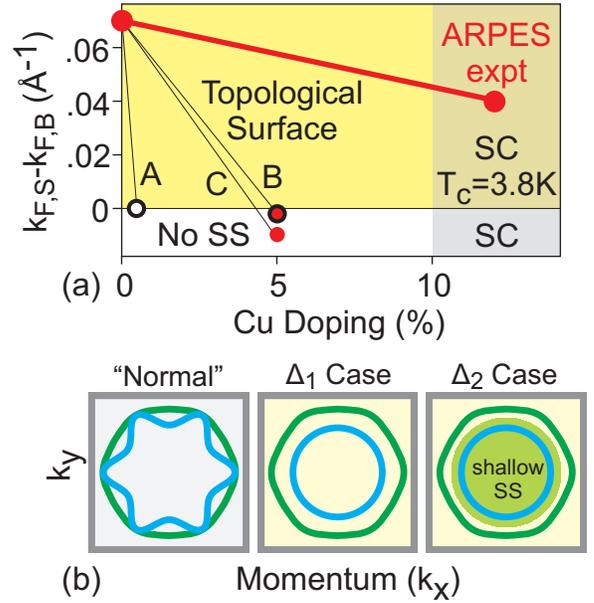}
\caption{\label{fig:ARPESphaseDiag}{\bf{Superconductivity on a topological surface}}: (a) A phase diagram compares (\textbf{ARPES expt}) the measured superconducting topology with preliminary expectations based on cases in which (\textbf{A}) each Cu atom donates one doped electron, (\textbf{B}) the experimental chemical potential is applied to GGA band structure, and (\textbf{C}) the experimental chemical potential is applied to the band structure of undoped Bi$_2$Se$_3$. (b) States within 5meV of the Fermi level are shaded for the k$_z$=0 plane of (center) a fully gapped doped topological insulator and (right) one example of a topological superconductor. A panel at left shows a scenario in which degeneracy between the bulk and surface electrons allows superconductivity to be achieved \emph{without} novel topological properties.}
\end{figure}

\begin{figure*}[t]
\includegraphics[width = 12cm]{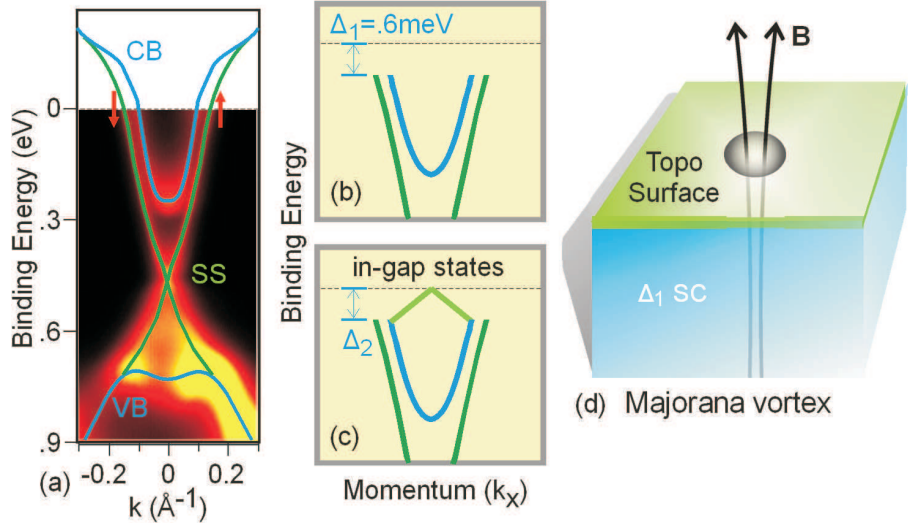}
\caption{\label{fig:platform}{\bf{A Majorana platform}}: (a) Topologically protected surface states cross the Fermi level before merging with the bulk valence and conduction bands in a lightly doped topological insulator. (b) If the superconducting wavefunction has even parity, the surface states will be gapped by the proximity effect, and vortices on the crystal surface will host braidable Majorana fermions. (c) If superconducting parity is odd, the material will be a so-called ``topological superconductor", and new states will appear below T$_C$ to span the bulk superconducting gap. (d) Majorana fermion surface vortices are found at the end of bulk vortex lines and could be manipulated for quantum computation if superconducting pairing is even ($\Delta_1$).}
\end{figure*}

Our experimental observations have several important implications for new superconducting states in Cu$_x$Bi$_2$Se$_3$.
First, our data show that the spin-textured surface and doubly degenerate bulk bands are well separated. This establishes that bulk superconducting pairing occurs in the presence of a non-degenerate, spin polarized two dimensional topological surface state. As a result, the proximity effect with the bulk superconductor is sufficient to induce the pairing of surface states, without the need of using an external superconductor as proposed earlier \cite{FuSCproximity}. The resulting 2D superconducting state on the surface is highly nontrivial: it is fully gapped but supports zero-energy Majorana bound states localized in the vortex core. These Majorana states have non-Abelian statistics and are potential building blocks of a topological quantum computer. For this reason, we call the superconducting state on the surface ``a 2D non-Abelian superconductor". Moreover, since Cu$_x$Bi$_2$Se$_3$ is
a type-II superconductor \cite{HorSC,HorAllSC}, the required vortex on the surface naturally appears at the end of a vortex line in the bulk \cite{Teo3D,FuVort}, and can be easily generated by applying a magnetic field (Fig. \ref{fig:platform}(d)).


Theoretical studies suggest that the number of Majorana fermions at the end of a vortex line is partly derived from bulk band structure properties at the Fermi level \cite{FuVort,HosurVort}. If the Fermi level is inside or very close to the bulk band gap of a topological insulator, then an odd number of Majoran fermions (e.g. one) is expected be bound to each end of a vortex line. If the Fermi energy is raised far above the conduction band minimum of the topologically inverted band gap, the 1D vortex line will eventually cross a topological phase boundary beyond which the end of each vortex will bind an even number of surface Majorana modes (e.g. zero or two). A recent study based on DFT band structures has predicted that the transition will occur when the Fermi level is raised higher than 0.24 eV above the conduction band minimum \cite{HosurVort}; however, the factor of $\sim$2 discrepancy we have observed between calculated DFT band kinetics and experimentally measured band structures in Cu$_x$Bi$_2$Se$_3$ and Bi$_2$Te$_3$ implies that the transition occurs at a higher energy, possibly E$_M$$\sim$2$\times$0.24 eV $\sim$0.5 eV above the conduction band minimum in Cu$_x$Bi$_2$Se$_3$. Because we observe the Fermi level to be only $\sim$0.25 eV above the conduction band minimum for optimally doped superconducting Cu$_{0.12}$Bi$_2$Se$_3$, the expectation based on our measurements of bulk band structure is that only one Majorana mode will be found at the end of each vortex. The well defined surface state observed in the measurements presented here is expected to give rise to a single Majorana mode that is localized within several nanometers of the crystal surface \cite{FuSCproximity}, and our ARPES measurements clearly rule out the possibility of a second Majorana bound state arising from \emph{surface states}. If a second mode exists (contrary to present predictions), it must come from bulk electrons, and therefore will be extended far more deeply along the vortex line into the material. In either case, the observation of non-Abelian phenomena made possible by the Majorana modes will depend on physical parameters that have yet to be investigated, such as detailed interactions and tunneling effects between vortices.



In addition to these surface properties, our data also provide important clues for understanding superconductivity in the bulk of Cu$_x$Bi$_2$Se$_3$. Strong spin-orbit coupling in the form of 3D massive Dirac bands plays a key role in the pairing symmetry of Cu$_x$Bi$_2$Se$_3$. As Ref.\cite{FuNew} shows, an unconventional spin-triplet pairing whose order parameter is odd under spatial inversion strongly competes with the conventional spin-singlet pairing. Specifically, two phenomenological parameters $U$ and $V$ were introduced in the two-orbital model, which denote the intra- and inter-orbital electron interactions respectively. It was found that triplet pairing wins at small electron density, provided that inter-orbital attraction exceeds over the intra-orbital one (see phase diagram in Fig. \ref{fig:FuPhaseDiag}). In reality, this condition is reasonable because larger intra-orbital repulsion tends to significantly weaken the phonon-mediated attraction.

The key band structure parameter constraining the form of superconductivity in the perturbative model presented in Section II is the ratio of rest mass (band curvature) to the doped chemical potential ($m/\mu$). In an ideal Dirac band structure described by the Hamiltonian in Eq. (1-2), the rest mass is equal to half of the band gap between the conduction and valence bands, or 150 meV. Because the observed chemical potential is 250 meV above the conduction band minimum, this gives a ratio of $\frac{m}{\mu}\sim\frac{1}{3}$ (blue arrow in Fig. \ref{fig:FuPhaseDiag}), which is small enough to allow the rare theoretical possibility that the bulk superconducting wavefunction will have an odd (-1) parity symmetry value. However, the warped shape of the valence band (partical-hole asymmetry) that we have discussed above causes different mass terms to be needed to describe the conduction and valence band lineshapes, with a larger mass term needed for the conduction band. For the experimentally derived fit parameters plotted in Fig. \ref{fig:DiracFit}, the ratio is raised to $\frac{m}{\mu}\sim\frac{3}{4}$, which would suppress the likelihood of odd superconducting pairing. The presence of a ``kink" in the dispersion centered near 90 meV binding energy causes a more linear dispersion locally at the Fermi level (see Fig. \ref{fig:DiracFit}(c)), making the velocity of Cooper-paired electrons consistent with the lighter band-gap rest mass and suggesting that the ratio of $\frac{m}{\mu}\sim\frac{1}{3}$ best describes the superconducting electron kinetics. The ``mass gap" between the valence and conduction bands comes directly from spin orbit coupling interactions, and is therefore a direct indicator of how orbital symmetries are influenced by spin-orbit coupling. Different detailed models may shift the boundary in the Fig. \ref{fig:FuPhaseDiag} phase diagram, but are not expected to change the overall physical scenario in which both even and odd states are possible for differently tuned pairing interaction strengths.

As a result, triplet pairing wins for a large and reasonable range of interaction parameters. Interestingly, the triplet superconducting state realizes a novel time reversal invariant ``topological superconductor'' phase, which is the electronic analog of the Balian–Werthamer (BW) phase in superfluid He-3. Recent theoretical research has shown that three dimensional topological insulators doped into a superconducting phase are classified by an integer topological invariant (``n"), just as the undoped topological insulator state is identified by a Z$_2$ invariant (parity invariant) \cite{FuNew}. The topological superconductor state expected under odd-parity pairing in Cu$_x$Bi$_2$Se$_3$ is identified by a non-zero value of ``n". A hallmark of this topological phase is gapless surface Andreev bound states within the pairing gap (see Fig. \ref{fig:ARPESphaseDiag}(b,right) and Fig. \ref{fig:platform}(c)). These Andreev bound states are unrelated to the electron surface states of the parent topological insulator Bi$_2$Se$_3$. The bound states host Bogoliubov quasi-particles which are two-dimensional Majorana fermions propagating on the surface. However, these delocalized Majorana modes are unlike the Majorana Fermions localized at vortices in the topological surface state, and no proposals currently exist for their use in quantum computing.

While triplet superconductivity has been postulated in several material systems (e.g. Sr$_2$RuO$_4$, UGe$_2$, ZrZn$_2$ and URhGe \cite{SrRuParity,SrRuReview,UGe_A,ItohUGe,ZrZn2,URhGe}), it is associated with the spontaneous breaking of time reversal symmetry in all of these cases due to the magnetic moment of triplet Cooper pairs. Breaking of time reversal symmetry rules out any three dimensional topological superconductor state \cite{SchnyderClass}. The spin-orbit coupled band structure of Cu$_x$Bi$_2$Se$_3$ allows a novel form of triplet Cooper pairing in which the total spin and orbital angular momentum is 0 (see Ref. \cite{FuNew} and the inset of Fig. \ref{fig:FuPhaseDiag}), unlike the known $p+ip$ pairing in which Cooper pairs have orbital angular momentum corresponding to a magnetic moment. The odd parity superconducting state suggested by the band parameters reported here for Cu$_{0.12}$Bi$_2$Se$_3$ is therefore non-magnetic, and the odd parity ``$\Delta_2$" superconducting state in the Fig. \ref{fig:ARPESphaseDiag} phase diagram is identified as a topological superconductor.

In summary, our measurements show that surface instabilities cause the spin-helical topological insulator band structure of Bi$_2$Se$_3$ to remain well defined and non-degenerate with bulk electronic states at the Fermi level of optimally doped superconducting Cu$_{0.12}$Bi$_2$Se$_3$, and that this is also likely to be the case for superconducting variants of p-type Bi$_2$Te$_3$. These surface states provide a highly unusual physical setting in which superconductivity cannot take a conventional form, and is expected to realize one of two novel states that have not been identified elsewhere in nature. If superconducting pairing has even parity, as is nearly universal among the known superconducting materials, the surface electrons will achieve a 2D non-Abelian superconductor state with non-commutative Majorana fermion vortices that can potentially be manipulated to store quantum information. Surface vortices will be found at the end of bulk vortex lines as drawn in Fig. \ref{fig:platform}(d). If superconducting pairing is odd, the resulting state is a novel state of matter known as a ``topological superconductor" with Bogoliubov surface quasi-particles present below the superconducting critical temperature of 3.8 $^o$K. As drawn in Fig. \ref{fig:platform}(c), these low temperature surface states would be gapless, likely making it impossible to adiabatically manipulate surface vortices for quantum computation. The unique physics and applications of the topological superconductor state are distinct from any known material system, and will be an exciting vista for theoretical and experimental exploration if they are achieved for the first time in Cu$_x$Bi$_2$Se$_3$.

Comprehensive spectroscopic images measured over a wide range of photon energies reveal the detailed massive Dirac-like bulk and surface conduction states and their topological structure at optimal superconducting composition, and reveal that the critical electronic parameter ratio $\frac{m}{\mu}$ is sufficiently large to render either parity state of superconductivity possible depending on the balance of Cooper pairing interactions present. These electronic and spin-orbital parameters provide the essential framework for device development based on the unique topological surface physics, and are an essential guide for future exploration of superconductivity in the strogly coupled spin-orbit systems of topological insulators.

\textbf{Acknowledgements:}

Authors acknowledge discussions with A. Kitaev, A. Ludwig, C.L. Kane, P. Lee and F.D.M. Haldane. The synchrotron X-ray-based measurements and theoretical computations are supported by the Basic Energy Sciences of the US DOE (DE-FG-02-05ER46200, AC03-76SF00098 and DE-FG02-07ER46352). Materials growth and characterization are supported by NSF/DMR-0819860 and NSF-DMR-1006492. M.Z.H. acknowledges additional support from the A.P. Sloan Foundation. L.F. is supported by the Harvard Society of Fellows (Harvard University) and thanks Erez Berg for collaboration. MZH acknowledges discussions with A. Vishwanath.

Note added: After the completion of this work we became aware of a report of superconducting Bi$_2$Te$_3$ under pressure \cite{pscBiTe}. The Bi$_2$Te$_3$ data reported here is relevant for ambient pressure superconductivity.

\end{document}